\documentclass[conference]{IEEEtran}

\usepackage{graphicx}
\usepackage[space]{grffile}
\usepackage{latexsym}
\usepackage{textcomp}
\usepackage{longtable}
\usepackage{multirow,booktabs}
\usepackage{amsfonts,amsmath,amssymb}
\usepackage{url}
\usepackage[utf8]{inputenc}
\usepackage[ngerman,english]{babel}

\usepackage{amssymb}

\usepackage{nicefrac}

\usepackage{amsmath}
\usepackage{amsfonts}

\newcommand{\calG}{\mathcal{G}}
\newcommand{\calN}{\mathcal{N}}
\newcommand{\calE}{\mathcal{E}}

\newcommand{\conftitle}{2016 IEEE/ACM International Conference on Advances in Social Networks Analysis and Mining (ASONAM)}

\usepackage{fancyhdr}
\fancypagestyle{pageStyleOne}{%
    \fancyhf{}
    \fancyhead[C]{\conftitle}
    \fancyfoot[C]{\thepage}    
}

\begin{document}

\title{Analyzing the Spread of Chagas Disease with Mobile Phone Data}

\author{
\IEEEauthorblockN{
Juan de Monasterio\IEEEauthorrefmark{1},
Alejo Salles\IEEEauthorrefmark{1}\IEEEauthorrefmark{2},
Carolina Lang\IEEEauthorrefmark{1},
Diego Weinberg\IEEEauthorrefmark{3}, 
Martin Minnoni\IEEEauthorrefmark{4}, 
Matias Travizano\IEEEauthorrefmark{4}, \\
Carlos Sarraute\IEEEauthorrefmark{4}
}
\IEEEauthorblockA{\IEEEauthorrefmark{1}FCEyN, Universidad de Buenos Aires, Argentina}
\IEEEauthorblockA{\IEEEauthorrefmark{2}Instituto de C\'alculo and CONICET, Argentina}
\IEEEauthorblockA{\IEEEauthorrefmark{3}Fundación Mundo Sano, Paraguay 1535, Buenos Aires, Argentina} 
\IEEEauthorblockA{\IEEEauthorrefmark{4}Grandata Labs, Bartolome Cruz 1818, Vicente Lopez, Argentina} 
}

\maketitle
\thispagestyle{pageStyleOne}
\pagestyle{fancy}

\begin{abstract}
We use mobile phone records for the analysis of mobility patterns and the detection of possible risk zones of Chagas disease in two Latin American countries. We show that geolocalized call records are rich in social and individual information, which can be used to infer whether an individual has lived in an endemic area. We present two case studies, in Argentina and in Mexico, using data provided by mobile phone companies from each country. The risk maps that we generate can be used by health campaign managers to target specific areas and allocate resources more effectively. 
\end{abstract}

\section{Introduction}

Chagas disease is a tropical parasitic epidemic of global reach, spread mostly across 21 Latin American countries. The World Health Organization (WHO) estimates more than six million infected people worldwide~\cite{who2016}.  Caused by the \textit{Trypanosoma cruzi} parasite, its transmission occurs mostly in the American endemic regions via the \textit{Triatoma infestans} insect family (also called ``kissing bug", and known by many local names such as ``vinchuca" in Argentina, Bolivia, Chile and Paraguay, and ``chinche" in Central America). In recent years and due to globalization and migrations, the disease has become a health issue in other continents~\cite{schmunis2010chagas}, 
particularly in countries who receive Latin American immigrants such as Spain~\cite{navarro2012chagas} and the United States~\cite{hotez2013unfolding}, 
making it a global health problem.

A crucial characteristic of the infection is that it may last 10 to 30 years in an individual without being detected~\cite{rassi2012american}, which greatly complicates effective detection and treatment. About 30\% of individuals with chronic Chagas disease will develop life-threatening cardiomyopathies or gastrointestinal disorders, whereas 
the remaining individuals will never develop symptoms.
Long-term human mobility (particularly seasonal and permanent rural-urban migration) thus plays a key role in the spread of the epidemic~\cite{briceno2009chagas}. Relevant routes of transmission also include blood transfusion, congenital contagion --with an estimated 14,000 newborns infected each year in the Americas~\cite{OPS2006chagas}--,
organ transplants, 
accidental ingestion of food contaminated by \textit{Trypanosoma cruzi}, and even, in a minor scale, 
by laboratory accidents.
The spatial dissemination of a congenitally transmitted disease sidesteps the available measures to control risk groups, and shows that individuals who have not been exposed to the disease vector should also be included in detection campaigns.

Mobile phone records contain information about the movements of large subsets of the population of a country, and make them very useful to understand the spreading dynamics of infectious diseases. They have been used to understand the diffusion of malaria in Kenya~\cite{wesolowski2012quantifying} and in Ivory Coast~\cite{enns2013human}, including the refining of infection models~\cite{chunara2013large}. The cited works on Ivory Coast were performed using the D4D (Data for Development) challenge datasets released in 2013. Tizzoni et al.~\cite{tizzoni2014use} compare different mobility models using theoretical approaches, available census data and models based on CDRs interactions to infer movements. They found that the models based on CDRs and mobility census data are highly correlated, illustrating their use as mobility proxies.

Mobile phone data has also been used to predict the geographic spread and timing of Dengue epidemics~\cite{wesolowski2015impact}. This analysis was performed for the country of Pakistan, which is representative of many countries on the verge of countrywide endemic dengue transmission. Other works directly study CDRs to characterize human mobility and other sociodemographic information. A complete survey of mobile traffic analysis articles may be found in~\cite{naboulsi2015mobile}, which also reviews additional studies based on the Ivory Coast dataset mentioned above.

In this work, we discuss the use of mobile phone records --also known as Call Detail Records (CDRs)-- for the analysis of mobility patterns and the detection of possible risk zones of Chagas disease in two Latin American countries. Key health expertise on the subject was provided by the \textit{Mundo Sano} Foundation.
 We generate predictions of population movements between different regions, providing a proxy for the epidemic spread. Our objective is to show that geolocalized call records are rich in social and individual information, which can be used to determine whether an individual has lived in an epidemic area. We present two case studies, in Argentina and in Mexico, using data provided by mobile phone companies from each country. 
 This is the first work that leverages mobile phone data to better understand the diffusion of the Chagas disease.

\section{Chagas Disease in Argentina and Mexico}

\subsection{Key Facts and Endemic Zone in Argentina}\label{endemic_zone_argentina}

\begin{figure}[h]
\centering
\includegraphics[width=0.85\columnwidth]{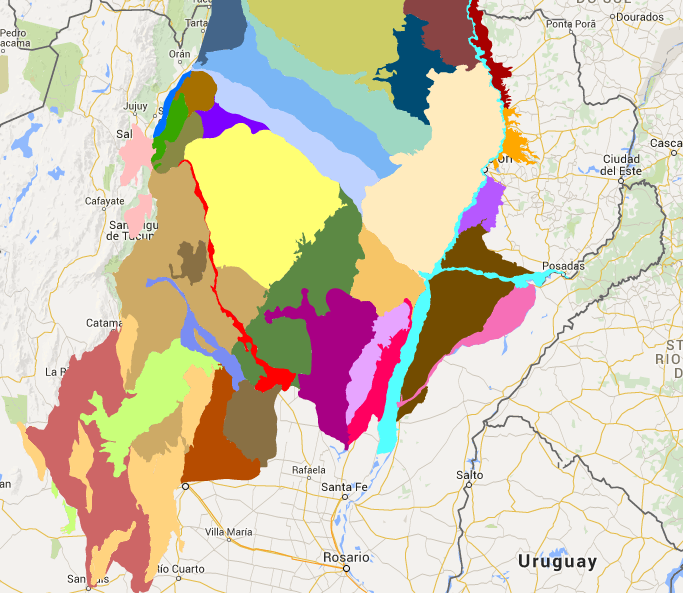}
\caption{The \textit{Gran Chaco} ecoregion in South America.%
}
\label{fig:granchaco}
\end{figure}

For more than 50 years, vector control campaigns have been underway in Argentina as the main epidemic counter-measure. The \textit{Gran Chaco}, situated in the northern part of the country,
is hyperendemic for the disease~\cite{OPS2014mapa}. A map of this ecoregion is shown in Figure~\ref{fig:granchaco}.
The ecoregion's low socio-demographic conditions further support the parasite's lifecycle, where domestic interactions between humans, triatomines and animals foster the appearance of new infection cases, particularly among rural and poor areas.
This region is considered as the endemic zone $E_Z$ in the analysis described in Section~\ref{methods} and Section~\ref{results}.

The dynamic interaction of the triatomine infested areas and the human mobility patterns create a difficult scenario to track down individuals or spots with high prevalence of infected people or transmission risk. Available methods of surveying the state of the Chagas disease in Argentina nowadays are limited to individual screenings of individuals. 
Recent national estimates indicate that there exist between 1.5 and 2 million individuals carrying the parasite, with more than seven million exposed. National health systems face many difficulties to effectively treat the disease. 
In Argentina, less than 1\% of infected people are diagnosed and treated 
(the same statistic holds at the world level).
Even though governmental programs have been ongoing for years now~\cite{plan_nacional_chagas}, data on the issue is scarce or hardly accessible. This presents a real obstacle to ongoing research and coordination efforts to tackle the disease in the region.

\subsection{Key Facts and Endemic Zone in Mexico} \label{endemic_zone_mexico}

\begin{figure}[h!]
\centering
\includegraphics[width=0.80\linewidth]
{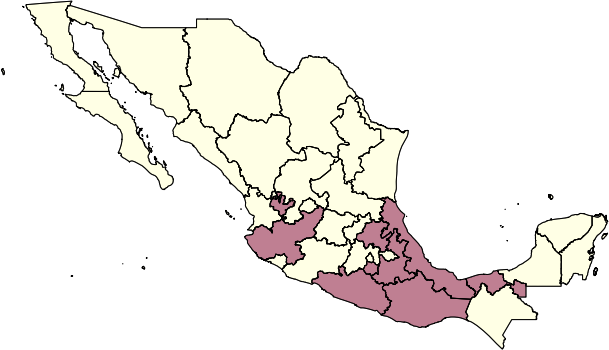}
\caption{Endemic region $E_Z$ for Mexico.}
\label{fig:endemic_zone_mexico}
\end{figure}

In 2004, the joint work of \textit{Instituto Nacional de Cardiología ``Ignacio Chávez"} and  \textit{Instituto de Biología de la UNAM} resulted in a Chagas disease database for Mexico~\cite{cruz2006chagmex}. Reviewing positive serology in blood banks and human reported cases per state, an epidemic risk map description was produced to geographically situate the disease. Based on this data, we defined the Mexican epidemic area, selecting the states having the top 25\% prevalence rates nationwide. The resulting risk region is shown in Figure~\ref{fig:endemic_zone_mexico}. It covers most of the South region of the country and includes the states of Jalisco, Oaxaca, Veracruz, Guerrero, Morelos, Puebla, Hidalgo and Tabasco.
This region is considered as the endemic zone $E_Z$ for the Mexican case in the analysis described in Sections~\ref{methods} and \ref{results}.

The authors of \cite{carabarin2013chagas} provide an extensive review of the 
research reports on Chagas disease in Mexico.
The review is very critical, stating that there are no effective vector control programs in Mexico;
and that the actual prevalence of the disease 
can only be estimated because no official reporting of cases is performed.

According to \cite{dumonteil1999update}, 
there are a total of 18 endemic areas in Mexico, located in the southeast, and
these areas include the states of Oaxaca, Jalisco, Yucatán, Chiapas, Veracruz,
Puebla, Guerrero, Hidalgo, and Morelos, all of them with rural areas.
Chiapas, Oaxaca, Puebla, Veracruz and Yucatán are among the most affected states (where the prevalence may exceed 10\%), although cases have been reported in most areas of the country
\cite{cruz2006chagmex,dumonteil1999update}.
Despite the lack of official reports, an estimate of the number of \textit{Trypanosoma cruzi} infections by state in the country
indicates that the number of potentially
affected people in Mexico is about 5.5 million~\cite{carabarin2013chagas}.
Mexico, together with Bolivia, Colombia, and Central
America, are among the countries most affected by this 
\textit{neglected tropical disease} (NTD)~\cite{hotez2013innovation}.
The disease doesn't know about borders:
Chagas and other neglected tropical diseases present in the north of Mexico remain highly endemic in the south of Texas as well~\cite{hotez2012texas}.

In recent years there has been a focus on treating the disease with two available
medications, benznidazole or nifurtimox. A study
that explores the access to these two drugs in Mexico 
shows that less than 0.5\% of those who are infected with
the disease received treatment in Mexico in years~\cite{manne2013barriers}.

People from endemic areas of Chagas disease tend to migrate to industrialized cities of the country, mainly Mexico City, in search of jobs. 
In accordance with this movement, a report showed
that infected children under 5 year of age are frequently distributed in urban
rather than in rural areas, indicating that the disease is becoming urbanized in
Mexico~\cite{guzman2001epidemiology}.
Therefore, as in the Argentinian case, the study of long-term mobility is crucial to understand the spread of the Chagas disease in Mexico.

\section{Mobile Phone Data Sources}

Our data source is anonymized traffic information from two mobile operators, in Argentina and in Mexico.
For our purposes, each record is represented as a tuple $\left < i, j, t, d, l \right >$,
where user $i$ is the caller, user $j$ is the callee, $t$ is the date and time of the call,
$d$ is the direction of the call (incoming or outgoing, with respect to the mobile operator client), and $l$ is the location of the tower that routed the communication.
The dataset does not include personal information from the users, such as name or phone number. The users privacy is assured by differentiating users by their hashed ID, with encryption keys managed exclusively by the telephone company.
Data was preprocessed excluding users whose monthly cellphone use either did not surpass a minimal number of calls $\mu$ or exceeded a maximal number $M$. This ensures we leave out outlying users such as call-centers or dead phones. In both datasets, we used $\mu = 5$ and $M = 400$.

We then aggregate the call records for a five 
month period into an edge list $(n_i, n_j, w_{i,j})$ where nodes $n_i$ and $n_j$ 
represent users $i$ and $j$ respectively and $w_{i,j}$ is a boolean value
indicating whether these two users have communicated at least once within the 
five month period. This edge list will represent our mobile graph  
$\calG = \left< \calN, \calE \right> $ where $\calN$ denotes the set of nodes (users) 
and $\calE$ the set of communication links. We note that only a subset $\calN_C$ nodes in $\calN$
are clients of the mobile operator, the remaining nodes $\calN \setminus \calN_C$ are
users that communicated with users in $ \calN_C $ but themselves are not clients of
the mobile operator. 
Since geolocation information is available only for users in $\calN_C$, in the analysis we considered the graph $\calG_C = \left< \calN_C, \calE_C \right> $ of communications between clients of the operator.

\paragraph{Datasets Information}
The Argentinian dataset contains CDRs collected over a period of 5 months, from November 2011 to March 2012. The raw data logs contain around 50 million calls per day.
The Mexican data source is an anonymized dataset from a national mobile phone operator. Data is available for every call made within a period of 19 months from January 2014 to December 2015. The raw logs contain about 12 million calls per day for more than 8 million users that accessed the telecommunication company's (\textit{telco}) network to place the call. This means that users from other companies are logged, as long as one of the users registering the call is a client of the operator. In practice, we only considered CDRs between users in $\calN_C$ since geolocalization was only possible for this group.

\paragraph{Data Limitations}
Although a lot of information is available in the CDR datasets, there may be limitations in their representativeness of the whole population. In each case, data is sourced from a single mobile phone operator, and no information is given on the distribution of its users, thus calls might in principle not represent correctly social interactions and movements between two given jurisdictions. Also, not all user movements will be captured by the log records. However, these limitations are offset by the huge datasets' sizes, from which we can safely assume that the amount of users observed in each set is sufficient to correlate real mobility or social links between different areas.

\section{Methodology} \label{methods}

In this section we describe the methodology used to generate risk maps for the Chagas disease in Argentina and in Mexico.

\subsection{Home Detection}

    The first step of the process involves determining the area in which each user lives. Having the granularity of the geolocated data at the antenna level, we can match each user $u \in \calN_C$ with its \textit{home antenna} $H_u$.  To do so, we assume $H_u$ as the antenna in which user $u$ spends most of the time during weekday nights. This, according to our categorization of types of days of the week, corresponds to Monday to Thursday nights, from 8pm to 6am of the following day. This was based on the assumption that on any given day, users will be located at home during night time~\cite{sarraute2015socialevents,csaji2012exploring}. 
   Note that users for which the inferred home antenna is located in the endemic zone $E_Z$ will be considered the set of \textit{residents of $E_Z$}.

In the case of Argentina, the risk area is the \textit{Gran Chaco} ecoregion, as described in Section~\ref{endemic_zone_argentina};
whereas in the case of Mexico, we used the region described in Section~\ref{endemic_zone_mexico}.

\subsection{Detection of Vulnerable Users}

    Given the set of inhabitants of the risk area, we want to find those with a high communication with residents of the endemic zone $E_Z$. To do this, we get the list of calls for each user and then determine the set of neighbors in the social graph $\calG_C$. For each resident of the endemic zone, we tag all his neighbors as potentially vulnerable. We also tag the calls to (from) a certain antenna from (to) residents of the endemic area $E_Z$ as \textit{vulnerable calls}.
    
    The next step is to aggregate this data for every antenna. Given an antenna $a$, we will have:
    \begin{itemize}
        \item The total number of residents $N_a$ (this is, the number of people for which that is their home antenna).
        \item The total number of residents which are vulnerable $V_a$.
		\item The total volume of outgoing calls $C_a$ from every antenna.
		\item From the outgoing calls, we extracted every call that had a user whose home is in the endemic area $E_Z$ as a receiver $VC_a$ (\textit{vulnerable calls}).
    \end{itemize}
    
    These four numbers $\left< N_a, V_a, C_a, VC_a \right>$ are the indicators for each antenna in the studied country.

\subsection{Heatmap Generation}
    With the collected data, we generated heatmaps to visualize the mentioned antenna indicators, overlapping these heatmaps with political maps of the region taken for study.

	A circle is generated for each antenna, where
		the \textbf{area} depends on the population living in the antenna $N_a$;
		and the \textbf{color} is related to the fraction ${V_a}/{N_a}$ of vulnerable users living there.
    
    We used two filtering parameters to control which antennas are plotted.
    \begin{itemize}
        \item $\beta$: The antenna is plotted if its fraction of vulnerable users is higher than $\beta$.
        \item $m_v$: The antenna is plotted if its population is bigger than $m_v$.
    \end{itemize}
    
    These parameters were tuned differently for different regions. For example: an antenna whose vulnerable percentage would be considered low at the national level can be locally high when zooming in at a more regional level.

\section{Results and Observations} \label{results}

\subsection{Risk Maps for Argentina}

\begin{figure}[h!]

\begin{minipage}{.495\linewidth}
\centering
  \includegraphics[width=0.90\linewidth]
  {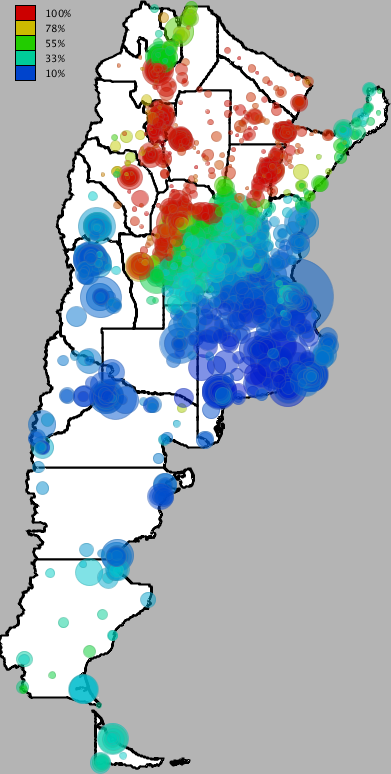}
  
(a) $\beta = 0.01$
\end{minipage}
\begin{minipage}{.495\linewidth}
\centering
  \includegraphics[width=0.90\linewidth]
  {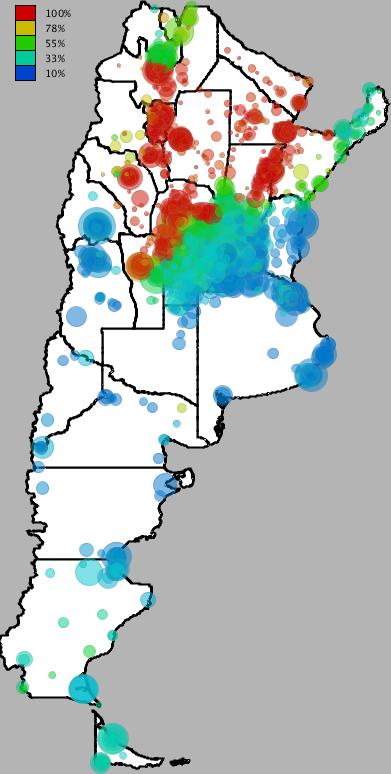}
  
(b) $\beta = 0.15$
\end{minipage}
\caption{Risk map for Argentina, filtered according to $\beta$.}
\label{fig:mapa_argentina}
\end{figure}

As a first visualization, maps were drawn using a provincial or national scale.
Advised by \textit{Mundo Sano} Foundation's experts, we then focused on areas of specific epidemic interest. 

Figure~\ref{fig:mapa_argentina} shows the risk maps for Argentina, generated with
two values for the $\beta$ filtering parameter, and fixing $m_v = 50$ inhabitants per antenna. After filtering with $\beta = 0.15$, we see that large portions of the country harbor potentially vulnerable individuals.
Namely, Figure~\ref{fig:mapa_argentina}(b) shows antennas where more that 15\% of the population has social ties with the endemic region $E_Z$.

\begin{figure}[h!]
\centering
  \includegraphics[width=0.95\linewidth]
  {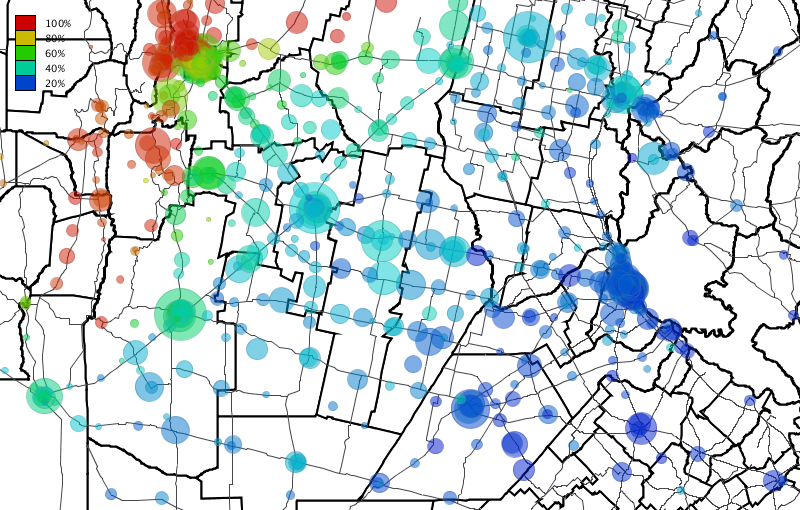}
\caption{Risk map for Cordoba and Santa Fe provinces, filtered according to $\beta = 0.15$.}
\label{fig:cba_sfe}
\end{figure}

Figure~\ref{fig:cba_sfe} shows a close-up for the Cordoba and Santa Fe provinces,
where we can see a gradient from the regions closer to the endemic zone $E_Z$ to the ones further away.

\subsection{Zooming and Detection of Vulnerable Communities}

\begin{figure}[h!]
\centering
\includegraphics[width=0.85\linewidth]{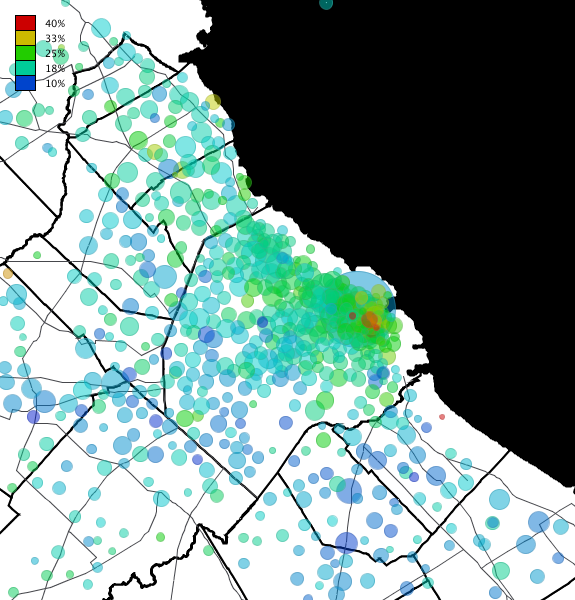}
\caption{Risk map for the metropolitan area of Buenos Aires, filtered with $\beta = 0.02$.}
\label{fig:amba_map}
\end{figure}

As a result of inspecting the maps in Figure~\ref{fig:mapa_argentina}, we decided to 
focus visualizations in areas whose results were unexpected to the epidemiological experts. 
Focused areas included the provinces of Tierra del Fuego, Chubut, Santa Cruz and Buenos Aires, with special focus on the metropolitan area of Greater Buenos Aires whose heatmap is shown in Figure~\ref{fig:amba_map}.

In some cases, antennas stood out for having a significantly higher link to the epidemic area than their adjacent antennas. Our objective here was to enhance the visualization in areas outside of Gran Chaco looking for possible host communities of migrants from the ecoregion.
High risk antennas were separately listed and manually located in political maps. This information was made available to the \textit{Mundo Sano} Foundation collaborators who used it as an aid for their campaign planning and as education for community health workers.

This analysis allowed us to specifically detect outlying communities in the focused regions. Some of these can be seen directly from the heatmap in Figure~\ref{fig:amba_map}, where the towns of Avellaneda, San Isidro and Parque Patricios have been pinpointed.

\subsection{Risk Maps for Mexico}

\begin{figure}[h!]
\begin{minipage}{.470\linewidth}
\centering
\includegraphics[width=0.95\columnwidth,height=2.3cm,keepaspectratio]
{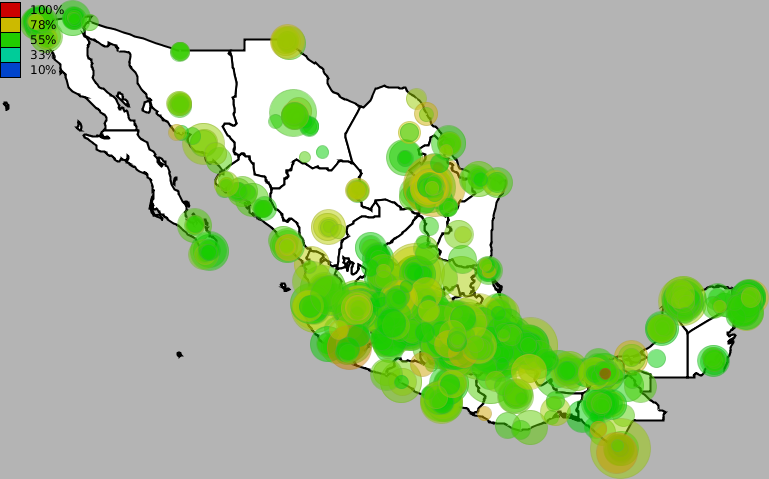}

\small (a) National map
\end{minipage}
\begin{minipage}{.520\linewidth}
\centering
\includegraphics[width=0.95\columnwidth,height=2.3cm,keepaspectratio]
{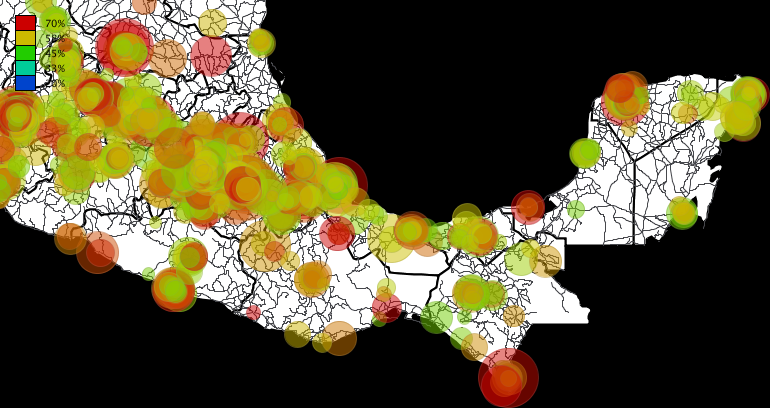}

\small (b) South region of Mexico
\end{minipage}
\caption{Risk maps for Mexico}
\label{fig:mapas_mexico}
\end{figure}

With the data provided by the CDRs and the endemic region defined in Section~\ref{endemic_zone_mexico}, heatmaps were generated for Mexico using the methods described in Section~\ref{methods}. The first generated visualizations are depicted in Figure~\ref{fig:mapas_mexico},
which includes a map of the country of Mexico, and a zoom-in on the South region of the country.
We used $m_v = 80$ inhabitants per antenna, and a high filtering value $\beta = 0.50$, which 
means that in all the antennas shown in Figure~\ref{fig:mapas_mexico},
more that 50\% of inhabitants have a social tie with the endemic region $E_Z$.
For space reasons, we don't provide here more specific visualizations and analysis of the regions of Mexico.

\section{Conclusions}

The heatmaps shown in Section~\ref{results} expose a ``temperature'' descent from the core regions outwards. The heat is concentrated in the ecoregion and gradually descends as we move further away. This expected behavior could be explained by the fact that calls are in general of a local nature and limited to 3 or 4 main antennas used per user. 

A more surprising fact is the finding of communities atypical to their neighboring region. They stand out for their strong communication ties with the studied region, showing significantly higher links of vulnerable communication. The detection of these antennas through the visualizations is of great value to health campaign managers. Tools that target specific areas help to prioritize resources and calls to action more effectively.

The results presented in this work show that it is possible to explore CDRs as a mean to tag human mobility. Combining social and geolocated information, the data at hand has been given a new innovative use different to the end for which the data was created (billing).

Epidemic counter-measures nowadays include setting national surveillance systems, vector-centered policy interventions and individual screenings of people. These measures require costly infrastructures to set up and be run. However, systems built on top of existing mobile networks would demand lower costs, taking advantage of the already available infrastructure. The potential value these results could add to health research is hereby exposed.

Finally, the results stand as a proof of concept which can be extended to other countries or to diseases with similar characteristics.

\section{Future Work}

The mobility and social information extracted from CDRs analysis has been shown to be of practical use for Chagas disease research. Helping to make data driven decisions which in turn is key to support epidemiological policy interventions in the region. For the purpose of continuing this line of work, the following is a list of possible extensions being considered:

\begin{itemize}
    \item \textit{Results validation.} Compare against actual serology or disease prevalence surveys. Data collected from fieldwork could be fed to the algorithm in order to supervise the learning. 
    \item \textit{Differentiating rural antennas from urban ones.} This is important as rural areas have conditions which are more vulnerable to the disease expansion. \textit{Trypanosoma cruzi} transmission is favored by rural housing materials and domestic animals contribute to complete the parasite's lifecycle. Antennas could be automatically tagged as rural by analyzing the differences between the spatial distribution of the antennas in each area. Another goal could be to identify precarious settlements within urban areas, with the help of census data sources.
    \item \textit{Seasonal migration analysis.} Experts from the \textit{Mundo Sano} Foundation underlined that many seasonal migrations occur in the \textit{Gran Chaco} region.  Workers might leave the endemic area for several months possibly introducing the parasite to foreign populations. The analysis of these movements can give information on which communities have a high influx of people from the endemic zone.
    \item \textit{Search for epidemiological data at a finer grain.} For instance, specific historical infection cases. Splitting the endemic region according to the infection rate in different areas, or considering particular infections.
\end{itemize}

\section*{Acknowledgements}

We thank Marcelo Paganini, Marcelo Abril and Silvia Gold from Fundación Mundo Sano for their valuable input, useful discussions and support of the project.
Special thanks to Adrián Paenza who provided the original idea and motivation of the project.

\IEEEtriggeratref{17}
\bibliographystyle{unsrt}
\bibliography{bibliography/epidemics}

\end{document}